\documentclass[12pt]{article}
\usepackage{graphicx}
\usepackage{xcolor}
\usepackage{amsmath,slashed, amssymb}
\usepackage{fancyhdr}
\usepackage{hyperref}
\usepackage{mathtools}
\usepackage{cite}
\usepackage[titletoc]{appendix}
\numberwithin{equation}{section} 

\def\be{\begin{equation}}
\def\ee{\end{equation}}
\def\beq{\begin{equation}\begin{aligned}}
\def\eeq{\end{aligned}\end{equation}}

\newcommand{\pd}[2]{\frac{\partial #1}{\partial #2}}

\newcommand{\dd}{\mathrm{d}}

\newcommand{\gsim}{\lower.7ex\hbox{$\;\stackrel{\textstyle>}{\sim}\;$}}
\newcommand{\lsim}{\lower.7ex\hbox{$\;\stackrel{\textstyle<}{\sim}\;$}}

\def\centeron#1#2{{\setbox0=\hbox{#1}\setbox1=\hbox{#2}\ifdim
\wd1\rangle\wd0\kern.5\wd1\kern-.5\wd0\fi
\copy0\kern-.5\wd0\kern-.5\wd1\copy1\ifdim\wd0\rangle\wd1
\kern.5\wd0\kern-.5\wd1\fi}}
\def\ltap{\;\centeron{\raise.35ex\hbox{$\langle$}}{\lower.65ex\hbox{$\sim$}}\;}
\def\gtap{\;\centeron{\raise.35ex\hbox{$\rangle$}}{\lower.65ex\hbox{$\sim$}}\;}
\def\gsim{\mathrel{\gtap}}
\def\lsim{\mathrel{\ltap}}

\newcommand{\newc}{\newcommand}
\newc{\qbar}{{\overline q}}
\newc{\Kahler}{Kahler }
\newc{\deltaGS}{\delta_{\rm GS}}

\begin{document}
\begin{titlepage}
\begin{flushright}
{\large SCIPP 20/01\\}
\end{flushright}

\vskip 1.2cm

\begin{center}

{\LARGE\bf Comments on Axions, Domain Walls, and Cosmic Strings}

\vskip 1.4cm

{\large Michael Dine$^{(a)}$, Nicolas Fernandez$^{(b,c)}$, Akshay Ghalsasi$^{(a,d)}$ and Hiren H. Patel$^{(a)}$}
\\
\vskip 0.4cm
{\it $^{(a)}$Santa Cruz Institute for Particle Physics and
\\ Department of Physics, University of California at Santa Cruz \\
     Santa Cruz, CA 95064.} \\
     ~\\
     {\it $^{(b)}$Department of Physics, University of Illinois at Urbana-Champaign \\
     Urbana, IL 61801.} \\
     ~\\
     {\it $^{(c)}$Illinois Center for Advanced Studies of the Universe \\
     University of Illinois at Urbana-Champaign \\
     Urbana, IL 61801.} \\
     ~\\
    {\it $^{(d)}$Department of Physics, University of Pittsburgh \\
     Pittsburgh, PA 15260.} \\
     ~\\
\vspace{0.3cm}

\end{center}

\vskip 4pt

\begin{abstract}
Axions have for some time been considered a plausible candidate for dark matter.  They can be produced through misalignment, but it has been argued that when inflation occurs before a Peccei-Quinn transition, appreciable production can result from cosmic strings.  This has been the subject of extensive simulations.  But there are reasons to be skeptical about the possible role of axion strings.  We review and elaborate on these questions, and argue that parametrically strings are already accounted for by the assumption of random misalignment angles. We review and elaborate on these questions, and provide several qualitative arguments that parametrically strings are already accounted for by the assumption of random misalignment angles. The arguments are base on considerations of the collective modes of the string solutions, on computations of axion radiation in particular models, and reviews of simulations.
\end{abstract}

\end{titlepage}

\section{Introduction and Summary of Main Results}

Axions have long been considered a promising dark matter candidate\cite{abbottsikivie,preskillwilczekwise,dinefischler}.  If the Peccei-Quinn transition occurs before inflation, they are produced by the so-called misalignment mechanism, with the result depending
on an essentially random parameter, $\theta_0$, the initial value of the angle $\theta$
within our present horizon.  If the
transition occurs after inflation, in a hot universe, then the misalignment mechanism still contributes, but the initial angle
varies by amounts of order $2\pi$ on Hubble scales. In this case, the dark matter density from misalignment is, in principle, computable, since one averages over the angles in different domains of the universe.

The plausibility of each of these scenarios depends on the value of $f_a$ and
the scale of reheating after inflation. Assuming the universe was once at temperatures higher than a few GeV, the axion decay constant, $f_a$,  must be small compared to, say, plausible scales of grand unification, and such scales {\it might} be associated with inflation (and more specifically reheating).  This might favor the post-inflation picture.  On the other hand, the scale of inflation (Hubble) can't be much higher than $10^{14}$ GeV \cite{Akrami:2018odb}, and it may not be easy to attain reheating temperatures above $10^{12}$ GeV. In any case, in this paper we will focus on the ``Post Inflation" scenario.

The texbtook \cite{kolbturner} estimate of the axion density assumes that the axion begins to oscillate around the minimum of its potential once $3 H \left( T_{\mathrm{osc}}\right) = m_a\left(T_{\mathrm{osc}}\right)$.  
The number density of axions is of order $\rho_a\left(T_{\mathrm{osc}}\right)/ m_a\left(T_{\mathrm{osc}}\right)$. So the axion energy density
behaves, for small initial misalignment angle, as:
\beq
\rho_a =\rho_a(T_{\mathrm{osc}}) \frac{m_a(T)}{m_a(T_{\mathrm{osc}})}  \frac{R^3(T_{\mathrm{osc}})}{R^3(T)} \,,
\eeq
where $R$ is the scale factor. In the case of PQ transition after inflation, one then averages over the initial misalignment angle, $\theta$, yielding $\theta^2 \rightarrow \frac{\pi^2}{3}$.

At the order one level there are several sources of uncertainty:
\begin{enumerate}
\item  Uncertainty as to the topological susceptibility, $\chi$,  the second derivative of the free energy at $\theta =0$:  $\chi$ is known analytically only at very low and very high temperature.   In the intermediate regime, one can estimate $\chi$ by interpolating between these results \cite{dinedraper,diCortona:2015ldu}, or one can attempt lattice simulations \cite{latticea,latticeb,latticec,latticed}.
\item Components of the axion field with Hubble scale variations are expected to contribute to the energy density an amount at least of order $f_a^2 H^2$, which is comparable to the assumed zero momentum
contribution. 
\item  Simply averaging the potential, proportional to $\sin^2\theta$, over $\theta$, does not take into account the non-linearity of the axion equations.  Here topological objects such as strings and domain walls, might enter. As with the previous item, these are likely to produce at least order one modifications of the axion density.   The question is
whether, for axions of momentum $H$ or smaller, these effects qualitatively alter the results obtained from consideration of the linear (misalignment) regime.
The claim in much of the literature on the subject is that these effects can, in fact, increase the axion dark matter density by orders of magnitude \cite{Davis:1986xc,Davis:1989nj,Battye:1993jv,Battye:1994au, dabholkar}. 
\end{enumerate}

We will argue in this paper that each of these uncertainties translate into order one (but not larger) uncertainties in the axion dark matter density. It would be desirable to reduce them, but their effect on estimates of the axion mass as a function of the dark matter density would seem modest \cite{dinedraper}. Our principle focus will be on the last item, and the possible role of cosmic strings (and at the final stages domain walls).  There is a large parameter in the problem, associated with infrared effects:
\beq
\xi = \log(f_a/H) \sim 70.
\eeq
The question then is:  is there a parametric enhancement of the axion dark matter density by powers of $\xi$?
A complete calculation requires numerical simulation; reliable analytic techniques are not available.  A variety of simulations exist, as we will review in section \ref{sec:simulations},
with somewhat disparate outcomes. To achieve this, the axion field surrounding the string must be converted entirely into low momentum axions.  Some simulations are consistent with this possibility, others not.  Much of this paper will be devoted to qualitative arguments that this does not occur.

There are several challenges to producing a sharp,
analytic argument for an enhancement by $\xi$. As is well known, the axion string tension is an infrared divergent quantity. As is typical for infrared divergences, this is an indication that the effective action should include, in addition to the string collective coordinates, light degrees of freedom (low momentum axions). Strings cannot be treated as entities unto themselves. Efforts to understand the possible role of cosmic strings analytically \cite{dabholkar,vilenkin} typically take as a starting point the Kalb-Ramond (KR) action for the coupling of the string degrees of freedom to the axion field \cite{kalbramond}. In some cases, these analyses yield an enhanced axion density.  We will explore analyses of this type. But treatment of the KR action requires some care.  We will give a partial treatment of the problem here, postponing a more detailed analysis for a subsequent publication.

There are several, qualitative reasons for skepticism as to enhancement by $\xi$, on which we will elaborate in this paper:
\begin{enumerate}
  \item  One can consider qualitative expectations for axion radiation for various string motions. In general, for momenta much higher than $H$, the system is adiabatic, and radiation of axions is suppressed.  For low momentum modes (of order $H$), if the motion is very slow, the system is again adiabatic and most of the energy stored in these long wavelength string modes is converted to string kinetic energy, suppressing low momentum axion production.  On the other hand, if the motion of these modes is fast, a sudden approximation is more appropriate.  The axion field circulating around the string is instantaneously converted to free axion radiation and this yields an axion  momentum spectrum behaving as $\log k$, where $k$ is the axion wave number, which does not {\it enhance} the low momentum axion density beyond the standard estimate. We might expect that except for non-relativistic motion of the strings, the sudden approximation would be closer to a true description of these low momentum modes; they ``will have a hard time keeping up" with the string. The sudden approximation leads to an axion energy density behaving as $d^3 k \, \rho \approx f_a^2 H^2 \frac{d^3 k}{k^3}$, where $k>H$.  This does not lead to an enhancement of the dark matter density. The real situation, particularly for the lowest momentum modes, is likely to lie in between these extremes, and we will that the results which are typical of these two limits do not support a parametric enhancement of the dark matter density.
     \item  The axion is a compact field.  As we will explain, this limits the amount of energy which can be stored in any particular range of axion wavelengths,
    and in particular in long wavelength modes (of order $k \sim H \sim 1/t$).  As we will explain, this is not simply a statement about precisely zero momentum axions,
    but about axions with a range of momenta near the Hubble scale.   One might wonder whether an enhanced string density (much larger than one per Hubble length) would evade this argument, but then, as we will see, production of axions from strings cuts off below
    momenta correspondingly larger than $H$.This ultimately tends to suppress the dark matter contribution.
\end{enumerate}

These considerations suggest a general framing of the question of possible
enhancement of the axion dark matter density:  allowing a modified distribution parametrized by a spectral index $q$,
\beq
\int d^3 k \, \rho(k,H) = f_a^2 H^2 \int \frac{d^3 k}{k^4}\left ( \frac{f_a}{k}\right )^{q-1}; ~H < \vert \vec k \vert < k_0 
\label{rhoofk}
\eeq
are there effects which lead to $q >1$, substantially increasing the numbers of low momentum axions?

The rest of this paper is organized as follows.  
In section \ref{axionstrings}, we review some aspects
of axion cosmic string solutions of the field equations, recalling the log divergence of the tension, and summarize the current status of numerical simulations. 
In section \ref{withoutstrings}, we discuss the evolution of an axion distribution with Hubble scale variation at the Peccei-Quinn phase transition (noting that there might not actually be a phase transition), ignoring non-linearities. We demonstrate that this leads to a universe with $q=1$.  We note that, assuming order one string per Hubble volume and a cutoff on the string width $\Lambda^{-1}$, we have a matching with the axion spectrum at scales below $\Lambda$.
In section \ref{simulations}, we review some results of simulations. In particular, in section \ref{sec:boundsonradiation}, we perform computations of the axion radiation using the critical string (KR) axion coupling. Setting any doubts about this procedure aside, we find that the resulting axion spectrum has $q\geq 1$, as we expect from other considerations.
We then discuss, in section \ref{collective},  collective coordinates for the strings and the separation of the axion excitations into low and high momentum modes, and the coupling of the low momentum modes to the string collective coordinates, remarking on the infrared divergence in the tension, aspects of the Kalb-Ramond action,
and the forces between segments of the string.  Our focus in this section is on non-relativistic motion, since this is the regime where field theory analysis of solitons is simplest, and we will be able
to learn some instructive, general lessons.  In general, we would expect, as supported by simulations, that the string motion is relativistic, and we will have in mind the passage from the non-relativistic to the relativistic regime.
In section \ref{adiabatic}, we lay out some naive expectations for radiation from axion strings, considering adiabatic and sudden limits. In section \ref{stringdensity} we discuss the possibility of a parametric enhancement of the density of strings, and its implications (or lack of them) for the axion density. In section \ref{domainwalls}, we discuss some aspects of domain wall production of axions. In section \ref{conclusions} we summarize, explaining why one does not expect an enhancement of low momentum axions due to cosmic strings, and the challenges to obtain an estimate better than order one of the axion dark matter density.

\section{Topological Defects:  Global Axion Strings}
\label{axionstrings}

In a post-inflationary scenario, a linearized treatment of the axion field is not reliable, as $a/f_a$ varies by $2\pi$ or more on Hubble scales.  Allowing such variation, one expects to encounter Hubble-scale closed loops for which
\beq
\oint \frac{a}{f_A}\,dl = 2 \pi n 
\eeq
for some integer $n$.  Indeed, shrinking the loop to smaller and smaller size, we must find regions in which the modulus of the underlying complex scalar field vanishes: cosmic strings.

In four dimensional field theories with spontaneously broken global $U(1)$ symmetries, there exist static solutions corresponding to infinitely long
strings.    The  tension of these solutions, however, is infrared divergent.  If $\Phi$ is the symmetry-breaking field, about a translationally invariant configuration, its fluctuations decompose into an axion field and a, generally heavy, fluctuation of the modulus:
\beq
\Phi = \left (f_a + \sigma(x) \right ) e^{i a(x)/f_a}.
\eeq
We can look for cosmic string solutions of the form, in cylindrical coordinates, $(\rho, \phi)$:
\beq
\Phi_{0} = f(\rho) e^{in \phi}.
\eeq
Necessarily $f(\rho)$ vanishes as $\rho \rightarrow 0$, while $f \rightarrow f_a$ as $\rho \rightarrow
\infty$.  More precisely,
\beq
f(\rho) \approx f_a \left(1 - \frac{\rm const.}{\rho^2 m_\rho^2}  \right).
\eeq
Away from the string core:
\beq
\vec \nabla \Phi_{0} = i \frac{n}{\rho} f_a e^{i n \phi} \hat \phi.
\eeq
From this expression, one sees that the energy stored in the string configuration per unit length is logarithmically divergent (we will see directly the divergence in the tension when we consider the collective coordinate analysis). The UV cutoff of the logarithmic divergence comes from the size of the string core.  The IR cutoff comes from physical considerations.  In practice, we do not expect the system to contain infinitely long strings.  Instead, we might imagine, for example, long, roughly parallel strings and anti{-}strings
(strings with opposite orientation of the $\phi$ variation); in this case the IR cutoff would be provided by the string separation. For closed, roughly circular, strings, the IR cutoff would be provided by the string circumference.  In a cosmic string network with $O(1)$ string per Hubble volume, we would expect the cutoff to be of order $H^{-1}$, and the string tension to be of order:
\beq
T = 2 \pi f_a^2 \log(f_a/H).
\eeq

\subsection{Numerical simulations}
\label{sec:simulations}
The possibility of substantial axion production by strings has been the subject of large scale numerical simulations\cite{villadoro,guymoore,Fleury:2015aca,sikivie1,sikivie2,sikivie3,Buschmann:2021sdq}. There is not complete agreement among the results. This is not surprising as calculating axions production by strings is the quintessential example of a physical problem with a vast separation of scales, $\log(f_a/H) \sim 70$.  Indeed, the issue raised by axion strings is whether there is an enhancement of the axion density by a power of this logarithm, corresponding to converting most of the energy in the axion field surrounding the string to low momentum axions.  Some recent simulations \cite{villadoro,guymoore,Vaquero:2018tib,Buschmann:2019icd,Gorghetto:2020qws} find that the relic abundance of axions from strings is smaller compared with the misalignment mechanism, but they disagree on the exact value.  Others \cite{Kawasaki:2018bzv,Hiramatsu:2010yu} find that the contrary is true: the contribution of the strings dominate over the misalignment mechanism. Additionally, while it has long been argued that the system reaches a scaling regime with a fixed average length of strings per Hubble patch, $\xi$, this has been called into question by a number of recent simulations \cite{Fleury:2015aca,Kawasaki:2018bzv,Klaer:2019fxc,villadoro,Vaquero:2018tib,Buschmann:2019icd,Gorghetto:2020qws,Buschmann:2021sdq}, observing a logarithmic growth\footnote{However recent results differ \cite{Hindmarsh:2019csc, Hindmarsh:2021vih, Hindmarsh:2021zkt}.}.  This suggests that $\xi$ may vary by $\mathcal{O}(10^2)$ factors at QCD temperatures between different simulations, which could translate to factors $\mathcal{O}(10)$ in the final abundance, if each string produced a $q=1$ spectrum, with lower momentum cutoff $H$. More important, the shape of the power spectrum of the axions is the crucial ingredient to determinate the axion number density and abundance. Depending of the spectral index, the power spectrum could be UV or IR dominated, with the axion distribution dominated by hard or soft momenta. Surprisingly, current simulations \cite{villadoro,Gorghetto:2020qws} claim evidence that the spectral index increases with simulation time, $f_{a}/H$, enhancing the number density of axions. Nonetheless, new results from a recent numerical simulation \cite{Buschmann:2021sdq}, using adaptive mesh refinement, disputes the increase of the spectral index and finds $q \approx 1$. Therefore, the diverse results and conclusions in the literature about the number density of axions from strings, could be explained by the extrapolation of $\xi$ and how the power spectrum is defined for extracting the number density.

\subsection{Skepticism as to an enhancement by powers of $\log(f_a/H)$}

In the rest of this note, we will argue by giving examples, that one does not expect an enhanced axion number density from defects such as strings and domain walls.  There will be several elements to the argument.
\begin{enumerate}
\item  Due to the infrared divergences, one must consider an effective action containing both string core collective coordinates and axions with momenta less than the inverse core size, where the cutoff length defining the core is chosen arbitrarily, but is short compared to the actual infrared cutoff on the system. We will perform the matching of the expected axion distributions at this cutoff.  We will give arguments which suggest, in support of some simulations, that the number of strings per Hubble volume might be parametrically larger than one, but explain that such an enhanced string density does not translate into an enhanced axion dark matter density.

\item  As discussed above, the logarithmic enhancement of the axion energy density from high momentum axions is also what is expected from considerations of Hubble scale variations of the phase of the Peccei-Quinn field, and their subsequent cosmic evolution. In this language, cosmic strings are just a piece of this set of variations. 
In particular, as we match the low momentum, axion component of the effective field theory with the string effective action at some cutoff, this matching is simple provided there is of order one string per Hubble volume.  As noted above, with more than one string per Hubble volume, we still expect a rough matching.
\item  For non-relativistic motion of the string, it is straightforward to derive an effective action for the string core collective coordinates and the axion.  This action indeed exhibits the cutoff-dependent tension.  Treated carefully, it also yields an axion-string coupling in this limit, which agrees with that for critical strings (Kalb-Ramond action).  Care is required in the treatment of
the axion background associated the string core.
\item  Similarly, in situations with a natural infrared cutoff, one can describe a priori expectations for axion radiation.  For example, in the case of a long parallel string and anti{-}string, one can think of the separation of the strings as a dynamical variable, $b(t)$, and obtain an action for $b$. One can demonstrate that, starting from rest and well separated, the system initially evolves in an adiabatic fashion.  This means that the potential energy of the separated strings is largely converted to kinetic energy of $b$; little is converted into low momentum axions at this early stage.  As the strings approach each other, a ``sudden" description becomes more appropriate, and the axion field of the system on scale $b(t)$ is converted into axions of wavelength $b^{-1}$, with energy $f_a^2 \ell$ ($\ell$ is the string length) per change of $b$ by a factor of $e$ (the base of natural logarithms).  
This is consistent with a contribution to the spectrum from strings with $q \le 1$, for small $k$, even with a possibly enhanced density of strings.  For an initially circular configuration, the sudden approximation is approached more rapidly, consistent with a distribution closer to $q=1$ even for small $k$.
\item  Even allowing for our skepticism of the conventional treatment of the Kalb-Ramond effect action to calculate
the axion emission spectrum, we perform the exercise in Sec.~\ref{sec:boundsonradiation} for various string motions using asymptotic analysis and find results consistent with $q \ge 1$ for large $k$.  However, the large $k$ part of the spectrum is dominated by the motion of the string near the end of the collapse, where the massive radial modes are more strongly coupled making  the Kalb-Ramond effective action, in any case, no longer appropriate.

\item  Most of the axion dark matter density is produced at later times, near the QCD phase
transition.  At these times, the axion field distribution between strings is, on scales of order
resemble strings, and one expects modes with distribution at low $k$ corresponding roughly to $q=1$.
\item  The axion is a compact field.  As a result, one might be suspicious of any
enhancement of the density at low momentum.

With $\Phi = f_a e^{i \theta(x)},$
if $\theta$ varies by $2 \pi $ on distance scales $\alpha^{-1} H^{-1}$, with $\alpha \ge 1$,  then the energy density, $\rho(x)$, is of order
\beq
\rho \sim \alpha^2 H^2 f_a^2.
\eeq
The corresponding momenta are of order $\alpha H$.
This suggests that any enhancement
of the energy density arises from population of higher momentum modes.  Modes with momentum much larger than $H$
make a suppressed contribution to the dark matter density.
If the axion energy density is to be enhanced by $\log(f_a/H)$, then one could have a dominant momentum of order $k \sim H \sqrt{\log(f_a/H)}$. Strings have $q=1$, and as a result, the {\it number density} of these higher momentum axions is suppressed.
\end{enumerate}

\section{Naive Expectations of the Axion Number Density from Hubble Scale Variations}
\label{withoutstrings}

A systematic treatment of the evolution of modes with general momentum is provided by working in conformal coordinates.  In terms of a reference time, $t_0$ (which we may
want to think of as the time of the PQ phase transition),
the scale factor satisfies, in a radiation dominated universe:
\beq
R(t) = \sqrt{\frac{t}{t_{0}}}R_0.
\eeq
If we define the conformal time, 
\beq
\tau = 2 \sqrt{t t_0}R_{0} = 2 t_0 R(\tau)
\eeq
then the metric is:
\beq
ds^2 = R(\tau)^2 (d\tau^2 - d\vec x^2).
\eeq
Let's turn to the case of the axion field in the regime where the potential is negligible. We can expand the $\phi(x,\tau)$ field
\beq
\phi(x,\tau) = \int \frac{d^3 k}{(2 \pi)^3} \phi(k,\tau) e^{i \vec k \cdot \vec x}.
\eeq
Now we can ask how $\phi(k,\tau)$ evolves with time.
In the conformal frame one has a simple equation:
\beq
\ddot \phi(k,\tau) + \frac{2}{\tau} \dot \phi + k^2 \phi =0.
\eeq
So for $k \ll \tau^{-1}$, i.e. modes outside Hubble horizon, $\phi(k,\tau)$ is essentially constant, and in general
\beq
\label{eq:kgtau}
\phi(k,\tau) = \phi(k, \tau_0) \frac{\tau_0}{\tau} \cos(k \tau).
\eeq

To compute the energy density we need to make some assumption about the mean value of $\phi(k, \tau_0)$.  Consider:
\beq
\langle \phi(k,\tau_0) \phi^\dagger(k^\prime,\tau_0)\rangle &= {\cal J}(\vec k,\vec k^\prime) \delta(\vec{k}-\vec{k^\prime}) &= \frac{f_a^2}{k^3} \delta(\vec{k}-\vec{k^\prime})
\label{kdistribution}
\eeq
where we have chosen a scale invariant spectrum
\beq
{\cal J}(\vec k,\vec k^\prime)  &=\frac{f_a^2}{k^3} \qquad  \mathrm{for} \qquad k \leq \frac{1}{\tau_{PQ}}
\eeq
and $\tau_{PQ}$ is the conformal time at PQ phase transition. Note that this choice gives us the correct energy density at the PQ phase transition. If we  consider the spatially averaged value of $(\vec \nabla \phi)^2$, we have:
\beq
\langle \partial_i \phi \partial_j \phi g^{ij} \rangle_{PQ}&= \frac{1}{R^{2}(t)} \int d^3 k  \frac{k^{2}}{(2 \pi)^3} \frac{f_a^2}{k^3} \theta(1/\tau_{PQ} - k) \\
&= \frac{f_a^2 t_0^2}{\tau^4} \propto f_a^2 H^{2}_{PQ}
\eeq
This is exactly as we expect.  Other choices of ${\cal J}$ would not behave correctly.

As the Hubble horizon expands more modes come within the horizon and begin to oscillate and decay according to Eq.~(\ref{eq:kgtau}). Now consider the contribution to the energy density
from larger momenta, from $k > \tau$.  Here, the modes start to oscillate for
$k = \tau(k)$, and damp as $\tau(k)^2/\tau^2 = \frac{1}{k^2} /\tau^2$.
So the corresponding contribution is:
\beq
\rho &= \left( \frac{1}{R(t)}\right)^{2}\int d^3 k  \frac{k}{(2 \pi)^3} \frac{f_a^2}{k^2} \frac{1}{\tau^{2} k^2} \theta(k-1/\tau ) \\
&= \frac{f_a^2}{\tau^2 R(t)^2} \log (\Lambda \tau) = {f_a^2 H^2} \log(\Lambda/H)
\eeq 
for some ultraviolet cutoff $\Lambda$.  A natural candidate for $\Lambda$ is $f_{a}$, 
the Peccei-Quinn scale. As we will see shortly, the higher momentum part of the spectrum, up to $f_a$, is filled in by strings.

In terms of the axion field, our analysis above translates to the correlation function (for times
corresponding to Hubble parameter $H$) as:
\beq
\langle a(\vec k) a(\vec k^\prime) \rangle = \frac{f_a^2 H^{2} }{ k^{5}} \delta(\vec k - \vec k^\prime) ~~~~
H< k < H_{PQ} 
\label{naiveaxioncorrelator}
\eeq
We will, in the next section compare this with expectations for strings, where we will see that the energy density stored in the strings is also enhanced by a $\log \left( \frac{f_{a}}{H} \right)$.  Note that in the case of strings the UV cutoff $\Lambda$ is $f_{a}$ and not $H_{PQ}$.

\begin{figure}
\centering
\includegraphics[width=0.65\textwidth]{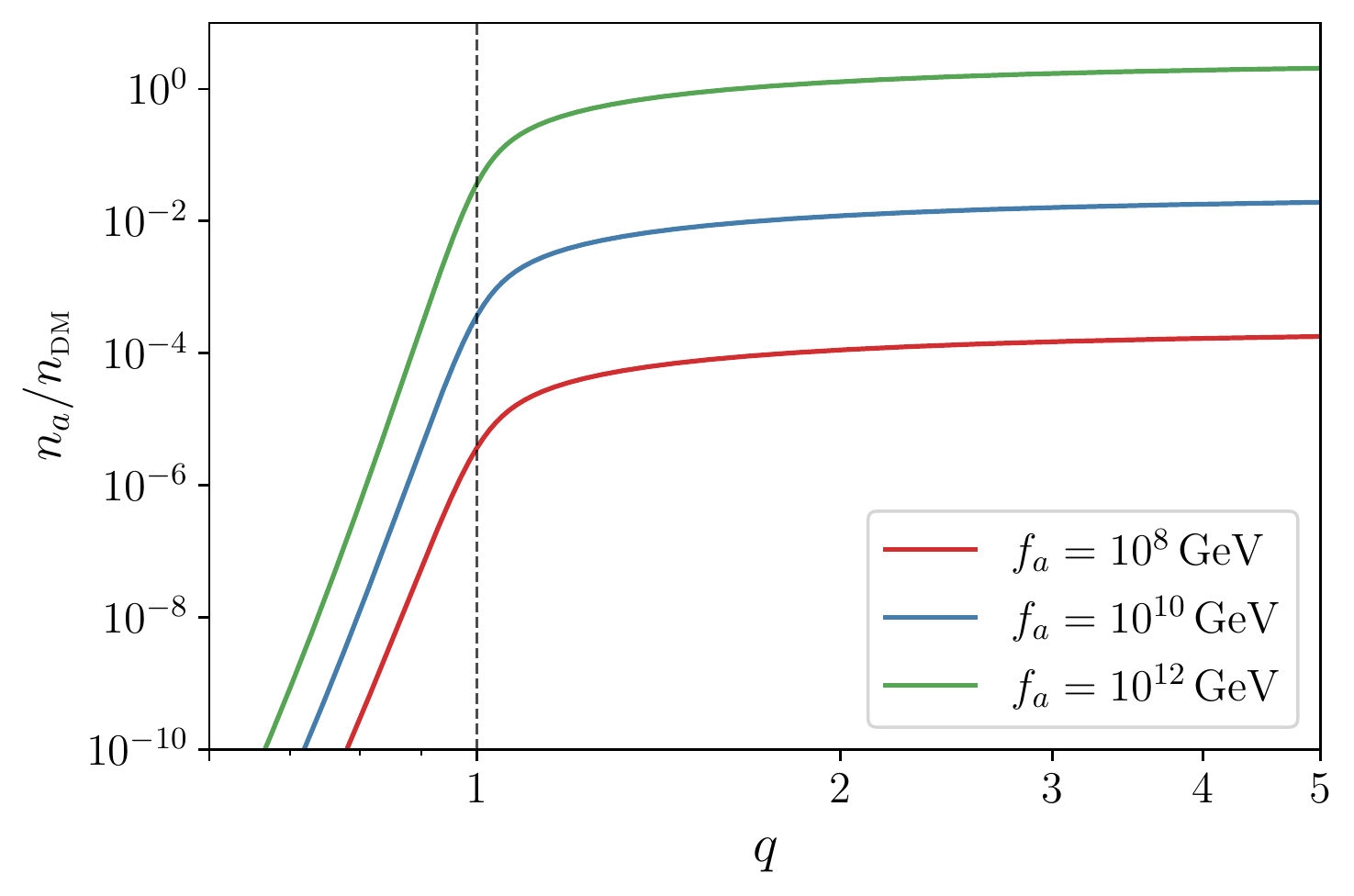}
\caption{The number density of axions as a function of spectral index for different values of  $f_{a}$ at QCD phase transition, normalized by the observed DM density. As can be seen the number density of axions, $n_{a}$, is a sharp function of spectral index if $q \lesssim 1$, while is soft once $q \gtrsim 1$.} 
\label{fig:DMvsq}
\end{figure}

The suggestion in the literature on cosmic strings is that, in fact, much of this higher momentum
scale axion distribution is converted into very low momentum axions.  Indeed, if the bulk of this energy were converted into low momentum axions, it is conceivable that these could
be the dominant source of axion dark matter\cite{hararisikivie,davisstrings,villadoro,guymoore}.
We can make the issue sharp by connecting the  spectral index, $q$, we defined earlier, to write the correlator of two axion fields, in
momentum space, as
\beq
\langle a(\vec k) a(-\vec k) \rangle = \frac{f_a^2 H^{q+1}}{k^{4+q}} \equiv \Delta(\vec k), ~  \vert k \vert \ge H.
\label{axioncorrelator}
\eeq
Then the axion energy density is:
\beq \label{eq:definitionOfq}
\rho =  \int \frac{dk\,}{2 \pi^2} \frac{d \rho(k)}{dk} \equiv \int dk \frac{f_a^2 H^2 H^{q-1}}{k^q}.
\eeq
$q=1$ corresponds to the axion field and energy distributions we have argued for above. Note that since axions gain mass after the QCD phase transition, we are concerned with the number density of the axions. In the string picture we can consider all the energy density stored in a string gets converted into axions. Axions from strings that decay early in the universe redshift significantly, but it can be shown that the principle contribution to the axions from decays of strings comes from the last Hubble time before the QCD phase transition. If we parametrize $\frac{d\rho}{dk} =\frac{C}{k^q}$ as above then we can find the normalization of the spectrum $C$ by equating the energy density from the spectrum to the energy density stored in the strings at QCD phase transition i.e. 
Then the axion energy density is:
\beq
\rho_{QCD} =  \int \frac{dk}{2 \pi^2} \frac{d \rho(k)}{dk}  \equiv \int dk \frac{C}{k^{q}} = f^{2}_{a}H^{2}_{QCD}\log\left(\frac{f_{a}}{H_{QCD}}\right) 
\eeq
The number density is then given by
\beq
n_{QCD} =  \int \frac{dk}{2 \pi^2} \frac{d \rho(k)}{dk} \equiv \int dk \frac{C}{k^{q+1}} 
\eeq
Then if $q > 1$, there is the potential for a significant contribution to the density at low momentum, as shown in Fig.~\ref{fig:DMvsq}.  Estimating the effects of decaying/annihilating strings on $q$ will be a principle focus of the remainder of the paper.

In any case, it should be stressed that $\log(f_a/H)$ is the large parameter which might
account for an enhancement of the dark matter density.  So the question is:  does the large logarithm in the tension can translate into a (parametric) logarithmic
enhancement of the axion dark matter density.

\section{Circular String Configurations and Their Collapse: Comparison with Simulations}
\label{simulations}

\begin{figure}[t]
\centering
\includegraphics[width=0.65\textwidth]{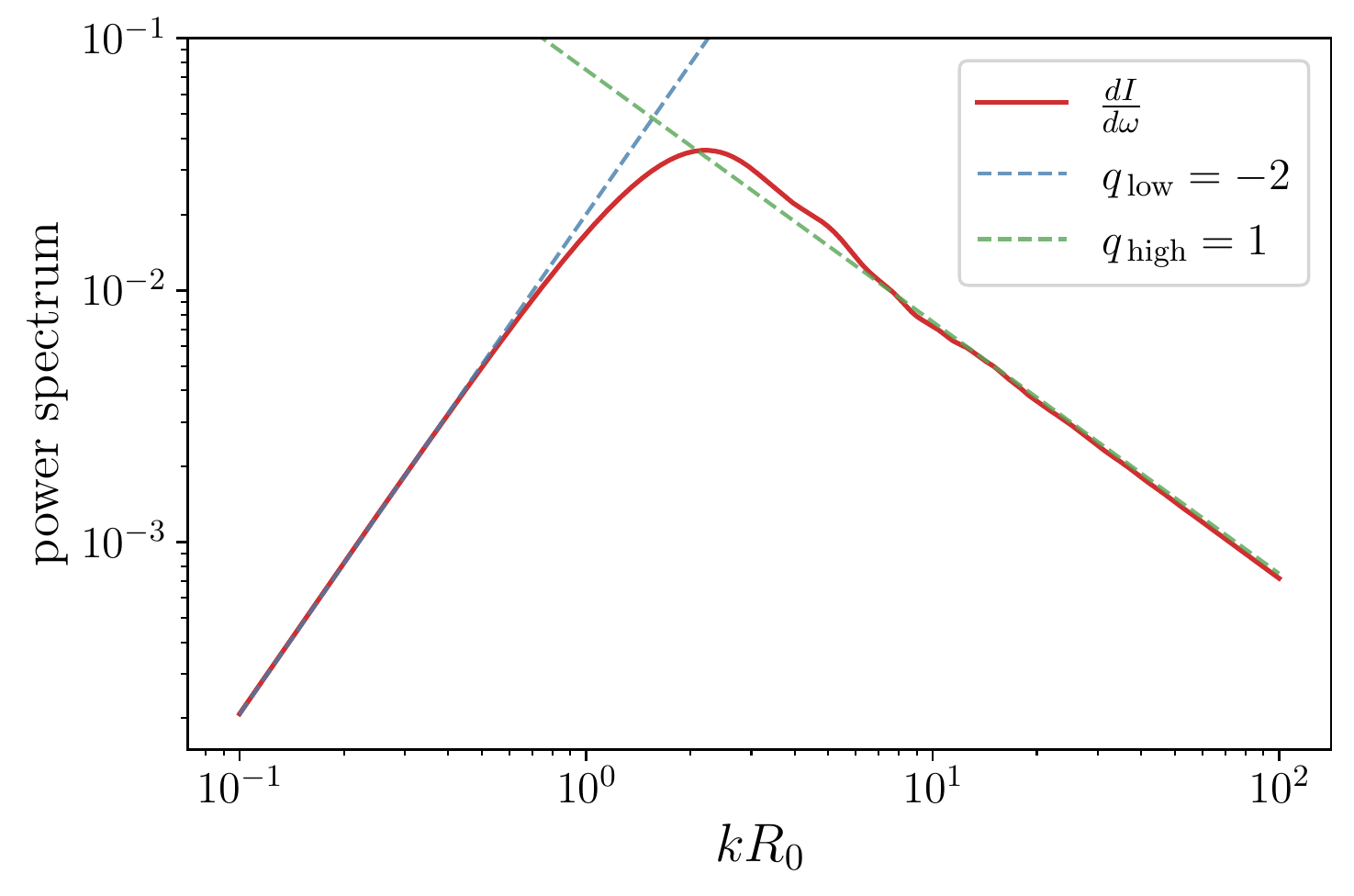}
\caption{Power spectrum of a circular string collapse as a function of frequency.} 
\label{fig:Power}
\end{figure}

Before considering collective coordinates and some aspects of radiation from non-relativistic strings, we review some of the arguments which have been given for substantial axion radiation from strings, and compare these with simulations.  We will use the frequently employed effective action in which the Nambu-Goto string couples to the Kalb{-}Ramond field summarized in Appendix A. For simplicity, we consider the idealized case of a perfectly circular string that is collapsing (Fig.~\ref{fig:Power}). We will see that such configurations have different features than those of long parallel strings.  In particular, if the string starts at rest, there is not a large inertia; the motion quickly becomes relativistic and the adiabatic approximation breaks down.

To see this, we will denote the string coordinates by $X=X^{1}(\sigma,\tau),~Y=X^{2}(\sigma,\tau)$, where, at time zero, the circular string of radius $R$ is in the $x$-$y$ plane,
\beq
X(\sigma) = R \cos(\sigma);~~~~Y(\sigma) = R\sin \sigma.
\eeq
Then the action for the collective mode, $R(t)$, for slow, non-relativistic motion is roughly:
\beq
S = \int dt f_a^2 R \log(f_a R) \left ( \dot R^2 -1 \right). 
\eeq
Starting from rest with $R = R_0$, energy conservation gives:
\beq
E = f_{a}^{2} R_0 \log(f_a R_0) = f_{a}^{2} R \log(f_a R) \left (\dot R^2 + 1 \right ).
\eeq
If $f_a R_0$ is very large, we can consider the case where $R = R_0 \rightarrow \frac{3R_{0}}{4}$, for example.  Then 
\beq
\dot R^2 = \frac{R_{0}}{R} -1 = 1/3,
\eeq
so the system quickly becomes relativistic.  We do not expect, in this case, that it is adiabatic for long.

We also expect that the circular geometries are favored, so that the system quickly collapses
and annihilates.  In the non-relativistic limit, it is easy to see that for closed strings moving in a plane and with fixed circumference, circular geometry minimizes (at least locally) the potential energy.
Note first that, from our collective coordinate analysis, calling $X(\sigma), Y(\sigma)$, the string coordinates, the potential energy of the configuration is:
\beq
V(X,Y) = \int_0^{2 \pi} d\sigma ( (X^\prime)^2 + (Y^\prime)^2) f_a^2 \log R f_a
\eeq
Here we are assuming that $f_a R \gg 1$, so small deviations in the length have little effect on the
log.
Writing
\beq
X(\sigma) = R \cos(\sigma) + \sum a_n e^{in\sigma}~~~Y(\sigma) = R \sin(\sigma) +\sum b_n e^{i n \sigma}.
\eeq
The constraint $\int ds \sqrt{\left( \frac{dX}{d\sigma} \right)^2 + \left( \frac{dY}{d\sigma} \right)^2}= R$, is for small $a_n$ and $b_n$, readily implemented.  The energy itself is quadratic in $a_n$'s and $b_n$'s so  is minimized, after enforcing the constraint, by setting all to zero (to get some feeling for the problem, consider first simply the case $a_n,b_n \ne 0$ for some fixed $n$). 

Once $R$ becomes of order $f_a^{-1}$, we expect production both of high momentum axions and of massive modes of the underlying theory. This picture finds some support in recent simulations \cite{Saurabh:2020pqe, Gorghetto:2020qws}.

\begin{enumerate}
\item  The initial configuration used in \cite{Saurabh:2020pqe} is not circularly symmetric.  The initial velocities are non-zero but are also non-relativistic.  In the early stages of the simulation, the system evolves to a closed circular loop, and the motion becomes relativistic.
\item  In \cite{Gorghetto:2020qws}, the circular string appears to collapse and annihilate in one or two oscillations.  On the other hand, in \cite{Saurabh:2020pqe} segments of the string break off as a result of collisions within the same string.  

\item The paper \cite{Saurabh:2020pqe} computes the spectrum at the end of the collapse of the circular string, and confirm that the spectral index of $q=1$ fits the data well.
\end{enumerate}

This seems quite close to the picture we have described above.  Indeed, the results of the simulations are consistent with $q=1$ for high momenta, with a fraction of energy in massive modes suppressed by $\log(f_a R_0)$.  These massive modes then convert to high momentum axions.

In the following subsection we make this intuition more concrete by doing an asymptotic analysis on the collapse of a closed string.

\subsection{Bounds on spectrum of radiated axions}
\label{sec:boundsonradiation}

The spectrum of radiated axions from the string can be written as $\frac{d\rho}{dk} \propto \frac{1}{k^{q}}$ as in (\ref{eq:definitionOfq}). Here $q$ is not necessarily a constant but can be a function of $k$ itself.  In this section we will show that there is a lower bound on $q$ for the collapse of a closed string. We will begin by analyzing the generic collapse of a perfectly circular string before generalizing it to an arbitrary closed string.

We can gain insight into the spectrum of radiated axions from a collapsing string by performing an asymptotic analysis on the radiated power and studying its $\omega = k$ dependence.  However this analysis is only applicable for $\omega \gg l^{-1}$ where $l$ is the relevant length scale of the string system. The analysis also relies on using the Kalb-Ramond action at the end of the life of the string, where the Kalb-Ramond description certainly fails \cite{dabholkar}.  

As shown in the Appendix, the radiation power from any string can be written
\begin{align}
\label{eq:rad}
\frac{dI}{d\omega} = \dfrac{\omega^{2} f_{a}^{2}}{16 \pi^3}\int d\Omega \left|\int d\sigma d\tau \left(\hat{n}\cdot(\vec{v}\times \vec{x}^{\prime})\right)e^{i\omega(x^{0}-\hat{n} \cdot \vec{x^{\prime}})}\right|^{2} \,,
\end{align}
where $x(\sigma,\tau) = (x^{0},\vec{x})$ is the four vector that describes the string, and $\sigma,\tau$ parametrizes the string world sheet. From now onwards we will drop the vector notation. $\vec{x}^\prime$ is the derivative of $\vec x$ w.r.t $\sigma$ while $\vec{v} = \dot{\vec{x}}$ is the derivative w.r.t $\tau$.
We can parametrize the circular string as $\vec x = R_{0}(f(\tau),\cos(\sigma)g(\tau),\sin(\sigma)g(\tau),0)$. Note that we have not used the solutions of E.O.M for a circular string and we keep the discussion general. Not having a particular solution allows us to ignore Hubble friction and feedback from radiation and is good enough to derive a lower bound on $q$. Plugging in the above parametrization of the circular string into (\ref{eq:rad}) we get
\begin{align}
\label{eq:rad1}
\frac{dI}{d\omega} \propto & \, \omega^{2} \int d\sigma d\sigma^{\prime} d\tau d\tau^{\prime} d\sin(\theta) \cos(\theta)^{2} g(\tau) g(\tau^{\prime}) \dot{g}(\tau) \dot{g}(\tau^{\prime})\\
& e^{i\omega R_{0}(f(\tau)-f(\tau^{\prime})+\sin(\theta)(\cos\sigma^{\prime}g(\tau^{\prime})-\cos\sigma g(\tau)))}
\propto \frac{1}{\omega^{q}} \,,
\nonumber
\end{align}
where we have redefined $\sigma - \phi$ as $\sigma$ and same for $\sigma^{\prime}$. In order to find the dependence of the radiated power on $\omega$ we will perform a stationary phase approximation  by assuming $\omega \gg R^{-1}_{0}$. For large $\omega$ the integrand will only have support when the derivative of the exponent vanishes. If we write the exponent as $$F(\tau ,\sigma ,\tau^{\prime},\sigma^{\prime}) = f(\tau)-f(\tau^{\prime})+\sin(\theta)(\cos\sigma^{\prime}g(\tau^{\prime})-\cos\sigma g(\tau)))$$ then the integral has support for parameters that satisfy  $\frac{dF}{d\tau^{(\prime)}} = \frac{dF}{d\sigma^{(\prime)}} = \frac{dF}{d\theta} = 0$. The solution to the set of equations is given by $\theta = \frac{\pi}{2},\sigma = \sigma^{\prime} = 0,\pi$ and some $\tau_{0}$ such that $g(\tau_0) = g(\tau^{\prime}_0) = 0$, \emph{i.e.} at the end of collapse of a circular string. Here we have also used the string gauge condition $\dot{x}^{2}+x^{\prime 2}=0$ which gives us $\dot{f}(\tau) = \sqrt{g^{2}(\tau)+\dot{g}^{2}(\tau)}$.

\begin{figure}
  \centering
  \includegraphics[width=.49\linewidth]{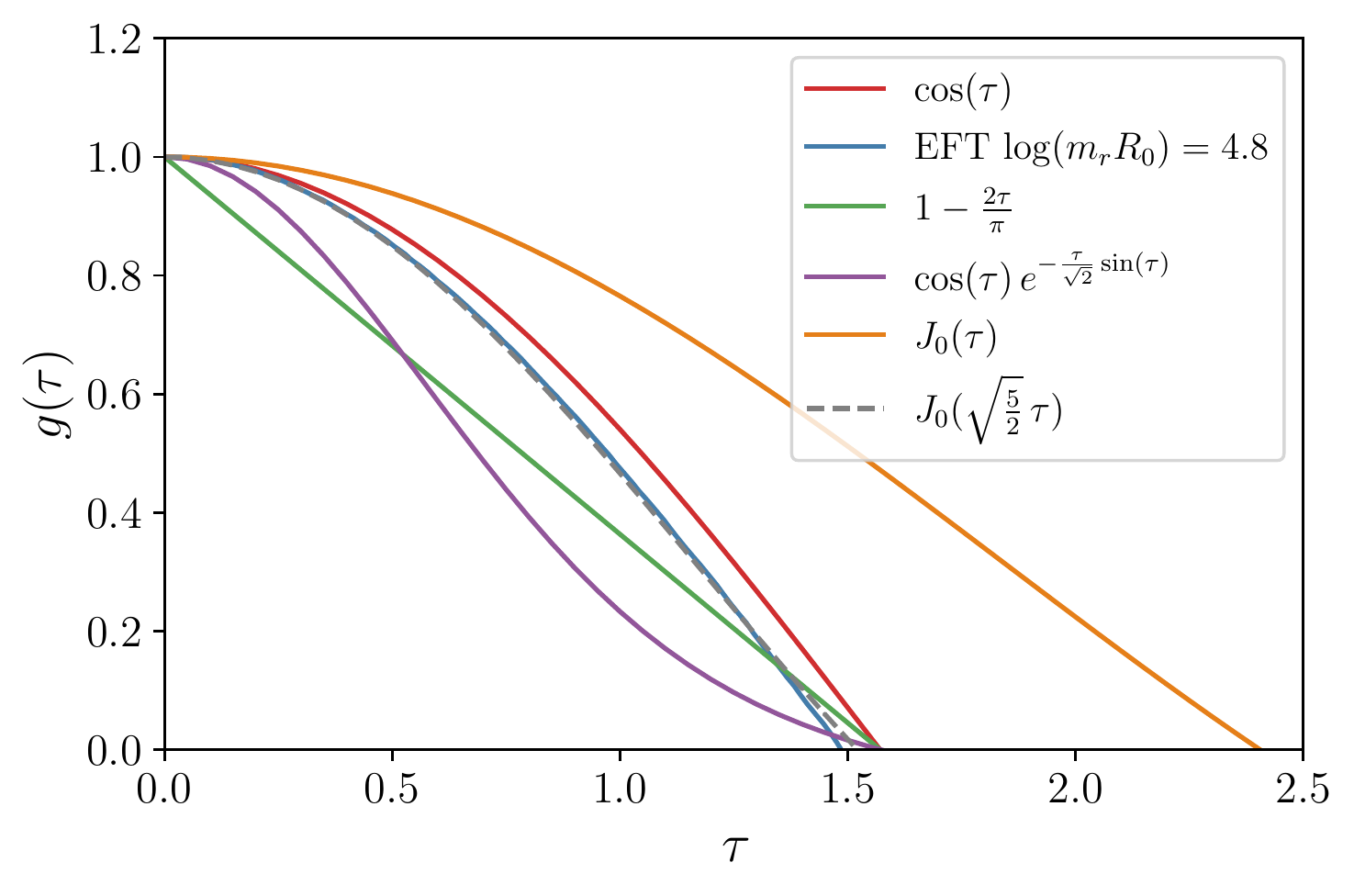}
  \includegraphics[width=.49\linewidth]{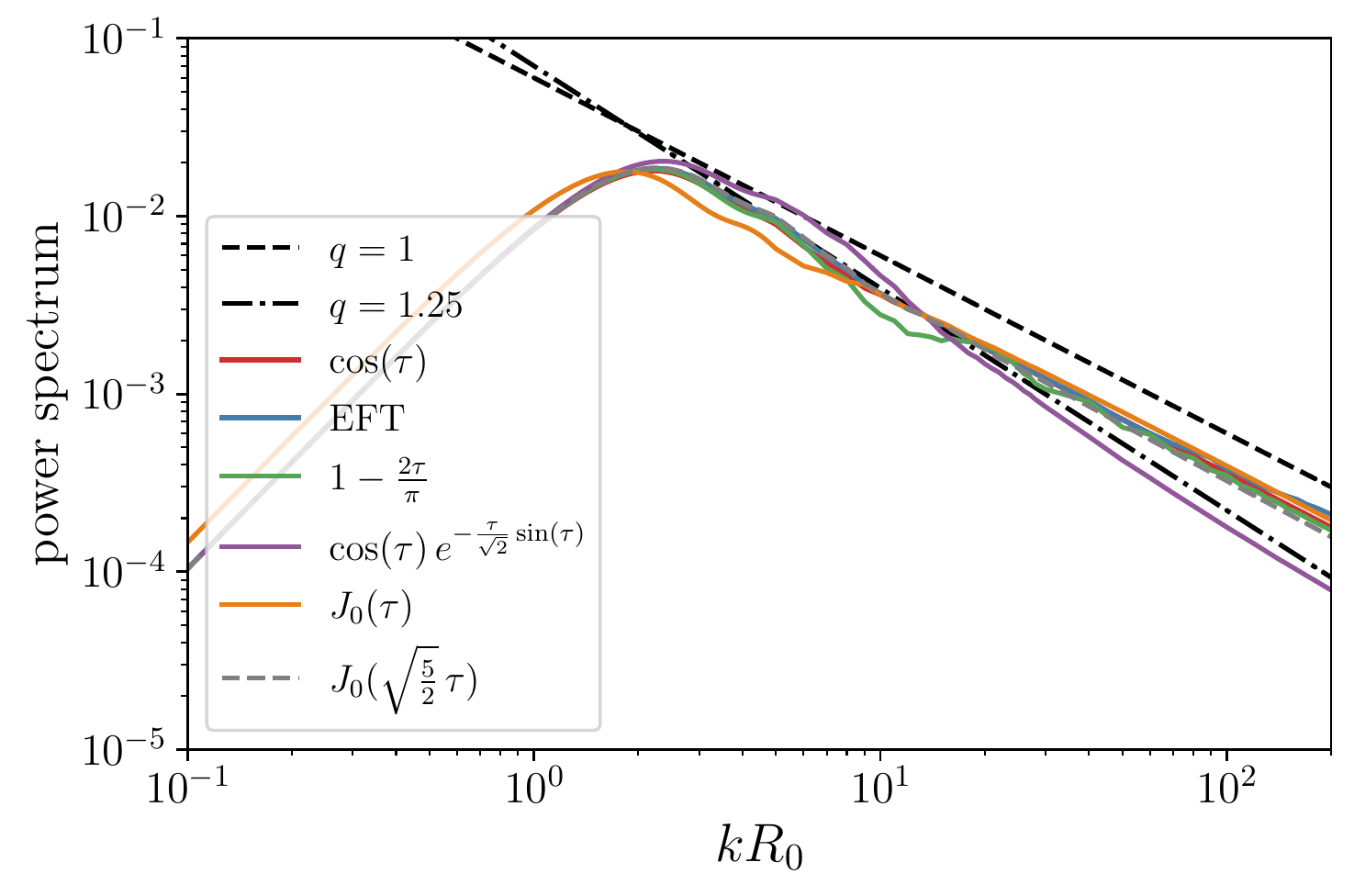}
\caption{Left panel shows the various motions of the string we consider whereas right panel shows the spectrum radiated from collapse of such a circular strings
}
\label{fig:diffgtau}
\end{figure}

We can then take $\tau_{0} = 0$ without loss of generality and Taylor expand all the terms around the respective extrema. If we write the Taylor expansion of the motion of the string 
\begin{align}
g(\tau) =  \sum_{\nu=n}^\infty \frac{\partial^\nu g}{\partial \tau^\nu}\Big|_{\tau=\tau_0}\frac{(\tau-\tau_0)^\nu}{\nu !} \,,  
\end{align}
where $n$ is the first non vanishing derivative of the motion of the string evaluated at the end of its motion. Taylor expanding the exponent, $F(\tau ,\sigma ,\tau^{\prime},\sigma^{\prime})$ and imposing the gauge condition we find that the first non-vanishing term in the exponent is the power of $n+2$ in the integrand variables $(\tau^{(')},\sigma^{(')},\theta)$. To remove the $\omega$ dependence in the exponent we rescale the integrand variables  i.e. $(\tau^{(')}, \sigma^{(')},\phi) \rightarrow \omega^{\frac{-1}{{(n+2)}}}(\tau^{(')}, \sigma^{(')}, \phi)$. We have five integration variables and expanding the non-exponent part of the integrand we find the first non-vanishing term is at the power of $4n$ in the integration variables. Thus we get that the radiated power using Eq.~(\ref{eq:rad1}) goes as $\frac{dI}{d\omega}\propto \omega^{2}\omega^{-\frac{4n+5}{n+2}}$.  This gives us the spectral index 
\begin{align}
\label{eq:boundcircular}
q = -2 + \frac{4n+5}{n+2} \geq 1, \qquad \omega \gg R_0^{-1} \,.
\end{align}
Thus for a circular string, we find that $q \geq 1$ always.  We emphasize that this result for the spectrum is valid for modes with frequency large compared to the inverse radius of the string.  Observe that the constant 5 in the numerator counts the number of independent integration variables. We have plotted the power spectrum for various string motions in Fig.~\ref{fig:diffgtau}. The asymptotic slope of radiated power always obeys $q \geq 1$.

Note that for a general string configurations, the loss of azimuthal symmetry prevents us from redefining $\sigma^{(')}- \phi$ as $\sigma$. This implies there are 6 independent integrations variables instead of the above 5 which makes the spectral index $q$ steeper.

\subsection{Numerical Kalb-Ramond Analysis for Circular Strings}

Above we considered a periodic $\cos(\tau)$ motion for the string collapse. Next, we will consider a variety of other motions to gauge the influence of the motion on the spectrum. The motions and their spectrum are plotted in Fig.~\ref{fig:diffgtau}. In choosing the motions we have kept the initial amplitude of the motion the same and all the motions are subluminal. Furthermore, we show the string motion calculated using the Kalb-Ramond action for radiating string interacting with its surrounding axionic field as show in \cite{dabholkar}. This back-reaction scenario is shown in Fig.~\ref{fig:diffgtau} under the label of EFT for $\log\left(m_{r} R_{0}\right)=4.8$\footnote{As $\log\left(m_{r} R_{0}\right)$ increases the circular string will tend to the $\cos(\tau)$ motion, i.e. the Nambu-Goto free string.}, and it is in good agreement with recent simulations (look Fig.~20 in \cite{Gorghetto:2020qws}).  
We then use Eq.~\ref{eq:rad} to numerically calculate the spectrum.  We find that most motions asymptotically have $q = 1$. The final number of axions radiated by the various motions is also fairly invariant with only few percent difference between the various motions.

We only consider circular strings for our calculation. So our results of $q = 1$ for circular motion is not proof that a string network with non-circular and non-planar string loops will produce a $q = 1$ spectrum for large $k$. Moreover, we only consider half a string motion at which point  we assume that the string annihilates. It's been pointed out \cite{dabholkar,Vilenkin:1986ku} that for $\log(f_{a}/H) \approx 70$, the string is underdamped and undergoes $\approx 20$ oscillations before it annihilates completely. Both non-circular and underdamped strings can give rise to a spectrum with $q > 1$. However as shown in \cite{dabholkar,Saurabh:2020pqe} (and as mentioned earlier in Sec. \ref{sec:simulations}), an initial configuration with kinks will quickly become circular and collapse. The underdamped oscillations in large $\log\left(f_{a}{H}\right)$ limit were derived using the Kalb-Ramond action which is expected to fail once the radius becomes smaller than $1/f_{a}$, where the production of the heavy radial modes occurs (again considered in \cite{Saurabh:2020pqe}).  Therefore, even though the string might undergo $\mathcal{O}(20)$ oscillations before annihilating, we restrict our analysis to a single oscillation to study the spectrum.
\section{The Effective Theory of Solitonic Strings}
\label{collective}

Infrared divergences generally signal that one must carefully include in the effective action low energy degrees of freedom.  In the present case, these are low momentum axions.  In this section, we formulate the problem of describing such
a system.  We first consider the case of gauged (Nielsen-Olsen) strings.  Here there are only
a limited set of low energy excitations, the collective coordinates of the solution, which can be fully accounted for by a Nambu-Goto like action.

\subsection{Collective Coordinates and Low
Momentum Axions}

Consider a long string in the Abelian Higgs model. Call $z$ the string direction; denote the transverse directions by $x$ and $y$. Take the classical configuration to be $\Phi_{0}(x,y)$ (constant in $z$, $t$). Now allow for a slow variation of the transverse coordinates with $z$ and $t$.  In other words, consider displacements, $X(z,t),~Y(z,t) \equiv X_{\bot}$, which vary slowly on the underlying scale of the theory (the masses of the charged scalar and gauge bosons). Substituting 
\beq \Phi(x,y,x,t) =  \Phi_{0}(x-X(z,t),y-Y(x,t))
\label{stringcollectivecoordinates}
\eeq
in the action, one obtains an action for the fields $X_0$, $Y_0$, from the $\partial_t,\partial_z$ terms in the action.  This is,
\beq
\int dt dz \left (\left (\frac{\partial \vec X_{\bot}}{\partial t} \right )^2-\left ( \frac{\partial \vec X_{ \bot}}{\partial z} \right )^2 \right )\int d^2 x_{\bot} (\partial_i \Phi_{0})^2 \,.
\eeq
The integral over the two transverse coordinates of the gradient of $\Phi$ (squared) is the same at all times and all $z$, due to the assumption of slow variation.  The integral over the
transverse coordinates yields the string tension.
This is the standard action for a non-relativistic string.

Note that it is important that the string is long.  In this case, the lowest ``stringy" modes have frequencies of order $1/\ell$, where $\ell$ is the length of the string.  They are thus separate from the excitations with frequencies typical of the mass scale of the underlying field theory.

Now contrast this with the axion string.  Here the problem is one of ``slow variation".  There are a continuum of axion excitations, extending down to momenta of order some low scale (typical string separations, or the radius of large closed strings).  No matter how large we take the scale of variation in the $z$ direction, the integral over the transverse coordinates is sensitive to this
cutoff.  If the cutoff is provided, for example, by a second, oppositely oriented string, the integral in the directions transverse to the
string is sensitive to the value of $z$; this is, as we expect, since the infrared cutoff is provided by the string separation, which varies with coordinate ($z$ or $\sigma$) along the string.  

So our low energy action should include both the excitations of the long string and axions with momenta below some cutoff $\Lambda$.
We could ask:  how might we choose $\Lambda$, and what does the theory with the cutoff look like?  In the cosmological setting, the cutoff, $\Lambda$, should be much larger than $H$.  Then the string tension is of order $f_a^2 \log(f_a/\Lambda)$, and it is necessary to consider axionic degrees of freedom with momenta lower than $\Lambda$ explicitly.  

\subsection{Radiation from Strings: Coupling of String Collective Modes to Axions}
\label{stringaxioncouplings}

We have seen that axion strings are significantly different than fundamental strings.  Because of infrared divergences, it does not make sense to speak of a single, long string in isolation. String-antistring configurations, or long closed strings, avoid this infrared problem, but collapse after a time scaled by their size or separation.  They leave over a spectrum of axions, so again it does not make
sense to speak of such strings in isolation.

The system may be described by an effective theory consisting of strings and axions,
where the string is essentially the core, out to some distance $\Lambda^{-1}\sim{f_a^{-1}}$, and includes the components of the axion field with momenta above a matching scale, $\Lambda$, and the axion field in the action contains
only momenta low compared to this scale.   A bit more precisely, we can think of a basis of axion wave packets.
We are interested in wave packets which scatter from the string, as opposed to those far
away.  It is these we are dividing into sets with typical momenta large or small compared to $\Lambda$.   

We can derive an expression for the coupling of the string collective modes to the
axion field, writing an expression similar to that of equation \ref{stringcollectivecoordinates}, allowing now both for small variations of the string in the transverse directions and for a spatially and time varying phase of $\Phi$. Again, the treatment parallels the  collective coordinate analysis for solitons such as monopoles.  Take the field, $\Phi$ (the complex PQ field) to have the form:
\beq
\Phi(z,t,\vec x_{\bot} )= \Phi_0(\vec x_\bot -X_{0\bot} )e^{i a(z,t,\vec x_\bot)/f_a}; ~~\vec x_\bot = (x,y) = (\rho,\phi).
\eeq
Plugging this into the action 
\beq
S= \int dz dt d^2x_{\bot} \partial_\mu \Phi^* \partial^\mu \Phi \, .
\eeq
We treat $\vec X_\bot$ as small and derivatives of $a$ as small, so we have, for example, from the
time derivative term:
\beq
\int d^2 x_\bot dz dt ~ \Phi_{0}^*~\partial_i \Phi_{0}  \dot X^i_\bot \partial_0 a/f_a + {\rm c.c.}; ~i=1,2.
\eeq
At large distances, far from the string core
\beq
\vec \nabla_\bot \Phi_0(\vec x_\bot) = f_a \vec \nabla_{\bot} e^{i\phi}  = i \hat \phi \frac{\Phi_{0}}{\rho} = i \partial_i \phi \Phi_{0} \, ,
\eeq
so the large distance part of the coupling is:
\beq
\int d^2 x_\bot dz dt ~f_a \partial_i \phi \partial_\alpha X^i_\bot \partial^\alpha a + {\rm c.c.};~\alpha=0,3.
\label{axionstringcoupling}
\eeq
A non-trivial coupling requires that the integral over $d^2 x_\bot$ be non-zero; this requires, in turn, that $a$ depend on the polar angle $\phi$.

\subsection{Comparison with the Kalb-Ramond action}
\label{Sec:Comparison}

In the literature, analytic attempts to estimate the energy radiated by strings\cite{vilenkin,dabholkar}
have taken cues from critical string theory.
In critical string theory, there is an axion, $a(x)$, dual to an antisymmetric tensor field, $B_{\mu \nu}$, through 
the relation
\beq
H_{\mu \nu \rho} = \epsilon_{\mu \nu \rho \sigma} \partial^\sigma a.
\eeq
Here $H_{\mu \nu \rho}$ is the field strength associated with $B_{\mu \nu}$, $H= dB$ in form
notation.  $B_{\mu \nu}$ couples to the string via:
\beq
\int d^2 \sigma  B_{\mu \nu} \frac{\partial X^\mu}{\partial \sigma_\alpha} \frac{\partial X^\nu}{\partial \sigma_\beta}\epsilon^{\alpha \beta}.
\label{kalbramondcoupling}
\eeq

Expression given in Eq.~\ref{kalbramondcoupling}  has been used to estimate radiation of axions from {\it axion strings}\cite{vilenkin,dabholkar}, though it is not a priori clear why 
the formula should apply to this case. Indeed, it is not at first sight obvious that our result, Eq.~\ref{axionstringcoupling}, matches the coupling of Eq.~\ref{kalbramondcoupling}.  
even in the extreme non-relativistic limit.

It is worth recalling the argument for the Kalb-Ramond coupling in the case of the critical string.  If a scalar in four dimensions is to be equivalent to an antisymmetric tensor field, it is necessary that the antisymmetric tensor be subject to a one-form gauge invariance, in order that there be only one physical degree of freedom:
\beq
B_{\mu \nu} \rightarrow B_{\mu \nu} + \partial_\mu \Lambda_\nu - \partial_\nu \Lambda_\mu \,.
\eeq
Then the three form, $H_{\mu \nu \rho}$ is gauge invariant:
$$H_{\mu \nu \lambda} = \partial_{[\mu} B_{\nu \rho ]}$$ (where the brackets indicate antisymmetrization of the indices). We require that the action should respect this invariance, as well as Lorentz invariance.  In addition to terms constructed from $H_{\mu \nu \rho}$, we also have the Kalb-Ramond term:
\beq
\int d^2 \sigma \frac{\partial X^\mu}{\partial \sigma_\alpha} 
\frac{\partial X^\nu}{\partial \sigma_\beta} \epsilon_{\alpha \beta} B_{\mu \nu}(X) \,.
\eeq
Under the gauge transformation, this goes to:
\beq
\delta S = \int d^2 \sigma \frac{\partial X^\mu}{\partial \sigma_\alpha} \frac{\partial X^\nu}{\partial \sigma_\beta}\epsilon_{\alpha \beta} 2 {\partial_\mu \Lambda_\mu}
=2\int d^2 \sigma \partial_\alpha \left (\epsilon^{\alpha \beta} 
\frac{\partial X^\mu}{\partial \sigma_\beta}\Lambda\right )\,.
\eeq
So the variation of $S$ is the integral of a total derivative, and vanishes for suitable boundary conditions on the string.

In attempting to apply this sort of reasoning to derive the effective action for the axion string,  a number of caveats are required.  One involves the use of four dimensional Lorentz invariance.
The axion solution, outside the core, as noted above, behaves as
\beq
\Phi = f_a e^{i \phi} \,,
\label{staticsolution}
\eeq
where $\phi$ is the angle of cylindrical coordinates.   
The four dimensional Lorentz symmetry of the underlying theory is broken by the string solution down to $SO(1,1) \times SO(2)$, where the $SO(2)$ is a combination of the original rotation symmetry and a Peccei-Quinn
transformation.  While the underlying theory may have the larger symmetry, it is not obvious how this should be realized in the effective Lagrangian, especially given that one has to include both string and low energy axion modes in this Lagrangian.

Correspondingly, it is necessary to define the radiation field relative to the axion field associated with the nearly static string.  In fact, carefully taking the difference of
the field $B_{\mu \nu}$ and the background $B_{\mu \nu}$ field associated with the static background, $\delta B_{\mu \nu}$, the source of $\delta B$ corresponds to
the source for the axion found in the collective coordinate analysis. It is critical to subtract this background configuration, and it is not entirely obvious how this should be
done in a relativistic setting.  Far away from a localized string, for specified motion, one can compute the radiation at long distances, as done in our Appendix B.  But this
has limitations in the cosmological setting, where the typical strings are of order horizon size.  In addition, the KR coupling generates, in the non-relativistic limit, an effective
potential for the string coordinates.  It is not immediately obvious how to generalize this to the relativistic case.  These issues will be discussed further in a subsequent publication.

In any case, we have used the KR action, in its conventional treatment, to discuss circular strings.  For very long parallel strings, or closed strings with long parallel segments,
a more complete treatment of the action would be required.  

\section{Expectations for Axion Radiation}
\label{adiabatic}

So far we have seen that global strings, unlike local ones, are entities which require careful definition.  One must introduce a cutoff at some length or momentum scale, and match onto the behavior of low momentum axions.  We have also seen that the standard Kalb{-}Ramond action does not describe the system, at least in the non-relativistic limit.  This leaves numerical simulations as the principle tool to investigate the list of order one uncertainties we have enumerated and, more urgently, whether there are parametrically large errors in the standard estimate.

In the effective theory description we have developed, however, there are reasons to expect that the resulting axion spectrum
has $q \le 1$.  We have alluded to two of these in the introduction.  The first comes from the fact that the axion is a compact field.  The second comes from consideration of two extreme limits of string motion:  adiabatic and sudden.

Let's consider, first, the issue of the axion as a compact field.
We can use this to bound the energy density as a function of
momentum, with $\Phi$ the PQ field, outside of the string core we have 
\beq
\Phi(x) = f_a e^{i \frac{a(x)}{f_{a}}} \equiv f_a e^{i \theta} \, .
\eeq
So we expect that the value of
\beq
\langle \partial_i \Phi \partial_j \Phi \rangle \le f_a^2 k^2 \, ,
\eeq
where $k$ is a momentum typical of the distribution.  $k \sim H$
is consistent with $2 \pi$ variation of $\theta$ in a Hubble
volume, or one string per Hubble volume. $k \sim {\cal N} H$ suggests
$2 \pi$ variation of the phase in a Hubble volume, for a larger number of
strings, scaling with a power of ${\cal N}$.  So, from this viewpoint, an enhanced axion density
is associated with a large number of strings, and a higher typical
axion momentum.  This is also consistent with the notion that if
there are many strings in a Hubble volume, the effective infrared cutoff on the string axion field configurations is larger by $a$, as the strings are packed closer together.

We have already mentioned the adiabatic and sudden approximations, but it is worth emphasizing what they suggest for axion radiation from the strings.  As the string moves through a Hubble volume with velocity some fraction of the speed of light, the axion component of the string, with $k \gg H$, will adjust instantaneously to the new string coordinates; there will be little radiation from these modes (suppression by powers of $k/H$).  However, modes with
momentum of order $k$ will not adjust to the new string coordinates, and some fraction will be shed. These will have momenta a spectrum, as we have already remarked, falling off as $1/k^3$.

\section{A Parametrically Enhanced String Density}
\label{stringdensity}

Some simulations have indicated the possibility that after some time, the system of axions plus strings evolves to an ``attractor" configuration with of order one
string per Hubble volume.  Other studies suggest a density which grows as $\xi =\log (f_a/H)$\cite{guymoore}.  Settling this question numerically
is challenging, since it is difficult to simulate the system at large values of the logarithm.  But our discussion of the
adiabatic and sudden approximations suggests that there might, indeed, be a logarithmic growth of the string density.
At the same time, our discussion of the effective theory indicates that such an enhancement {\it does not} translate
to an enhancement of the axion dark matter density.

The issue can be seen in our discussion of non-relativistic string motion.  For two long parallel strings, separated by a distance of order
$H$, and with opposite orientation, there is an attractive force, but there is also a large inertia -- enhanced by the same logarithm.
As a result, the time required for the two strings to approach and annihilate is decreased relative to $H$ by $\sqrt{\xi}$.
We might expect that if there is a sort of attractor solution with of order one string per Hubble volume, strings annihilate in a time of order a Hubble time.  Then requiring
that the string separation be suppressed by a compensating power of $\xi$.  Indeed, crudely, one expects a single power.  
consistent with some simulations.  But for this to be the dominant
effect requires that the string distribution more closely resembles
a collection of long parallel string pairs.  If the typical configuration
involves circular strings, one does not have a suppression of the
string velocity by the large inertia (tension), and one might not
expect a growth of the density.  We have already remarked that
if circular configurations dominate, one expects $q \approx 1$.

In any case, a parametrically enhanced density of strings does not translate to an enhanced dark matter density.  First, because the strings are packed
more closely together, say by $(\sqrt{\xi}H)^{-1}$, the effective infrared cutoff on the string solution occurs at larger momentum.
Similarly, in the non-relativistic picture, due to the large inertia, we would expect an adiabatic approximation to be valid,
and most of the energy of the collapsing strings would be converted to string kinetic energy, rather than axions, until they collide.
This remains true in the period of domain wall domination, described in the next section.

\section{Axion Domain Walls}
\label{domainwalls}

The dominant production of dark matter axions is associated with the latest times.  In practice, we would expect this
to mean times at which the axion mass is comparable to $H$.  At this time, if we consider a string-antistring pair of length of order $b= H^{-1}$
separated by a distance $b$, the energy of the configuration is of order $f_a^2 H^{-1} \log(f_a/H) +
m_a^2 f_a^2 H^{-3}$ arising from the $m_a^2 a^2$ term in the axion potential.  So it quickly becomes
energetically favorable (once $m_a \gg H$) to flatten out the axion configuration, i.e. to form domain wall
configurations.  These configurations are three dimensional analogs of the Sine-Gordon soliton in $1+1$
dimensions.  For string pairs separated in the $x$ direction, the typical domain wall configuration has the
form
\beq
a = 2 f_a \tan^{-1} (m_a x)
\eeq
with associated tension of order $f_a^2 m_a$.  The corresponding energy is 
\beq
m_a f_a^2 H^{-2} = m_a^2 f_a^2 H^{-3} \left (\frac{H}{m_a} \right ).
\eeq

For our string-antistring configuration, once $m_a \gg H$, there will be a strong force pulling the strings together
and leading to their annihilation.  If the strings are separated by a distance $x$, the potential $m_a f_a^2 H^{-1} b$
should be compared to $f_a^2 H^{-1} \log (f_a/H).$
From energetic configurations, we expect that every string or string segment will pair with a nearby anti-string.
The string network quickly disappears.

Returning to rather simple-minded estimates, we expect of order one domain wall per horizon.  The strings bounding the domain walls, for $k > H$, would be expected to resemble the corresponding string solutions. So, again, we would expect the dominant effects to be from axions with momenta of order $m_a$ or less, with a density of order $m_a^2 H^2 \sim H^4,$ i.e. an order one contribution to the standard estimate.

In the case of an enhanced string density, the domain walls form later, since the typical axion momentum is larger; this leads to a compensating suppression of the dark matter axion density.

\section{Conclusions}
 \label{conclusions}
 
 If axions play a role in the solution of the strong CP problem,
 there are several possibilities for the role of axions in the early universe.
 \begin{enumerate}
 \item  The Peccei-Quinn symmetry is broken before inflation.  In this case, the axion dark
 matter density depends on the axion decay constant and a random angle, $\theta_0$, known as the
 misalignment angle.  If the universe was in thermal equilibrium at temperatures well above the QCD scale, then there is a lower limit on $f_a$, of order $10^{12}$ GeV if  $\theta_0$ was of order one.  $f_a$ might be larger if $\theta_0$ was simply small, or was small for anthropic reasons \cite{lindeaxions}.
 \item  The explicit breaking of the PQ symmetry was large during inflation; in this case, $\theta_0$ is fixed, and presumably a number of order one \cite{anisimovdine}.  In this case, limits on $f_a$ are not likely relaxed by considerations of $\theta_0$.
 \item  Inflation ended after the PQ transition, but the universe only reheated to a very low 
 temperature.  In this case, the constraints on $f_a$ are significantly weakened, with
 $f_a$ somewhat smaller than $10^{15}$ GeV \cite{dinefischler,banksdinegraesser} allowed, even with $\theta_0 \sim 1$.
 \item  The PQ symmetry is spontaneously broken after inflation.  In this case, $\theta_0$ takes different values in horizon-size regions of the universe, and the axion dark matter density should be computable.  
 \end{enumerate}
 
 In this paper, we have focused on the last scenario.
 There has been a long running debate about whether cosmic strings play an important role in the determination of the axion density.  Because theses strings have a logarithmically enhanced tension, there is the possibility that they produce an enhanced contribution to the axion dark matter density. This would lead to a different relation, for example, between axion mass and dark matter density than otherwise.  Lattice simulations provide some evidence both for and against this possibility.
As we have stressed here, in order that these configurations be important, is it critical that they deposit, when they decay or annihilate, an order one fraction of their energy in Hubble-scale momentum. 

Before considering the strings themselves, we noted some general considerations about Hubble scale variation. In particular, one expects two components in the axion momentum distribution, at typical times above the QCD temperature:
\begin{enumerate}
\item  A low momentum component, with momentum of order $H$ or smaller, and energy density of order
$f_a^2 H^2$.
\item  Axions with momentum greater than $H$, and up to some cutoff, $\Lambda$, with distribution
\beq
\int d^3k \rho(k) = \int d^3 k f_a^2 H^2 \frac{d^3k}{k^2} \left ( \frac{f_a}{k} \right) ^q \Theta(\Lambda-k) \,.
\label{logdistribution}
\eeq

\end{enumerate}
We have argued that the compactness of the axion field bounds the first component of the density, above, at $f_a^2 H^2$.  By itself, this makes it hard to understand an enhancement of the
axion density over the standard computation \cite{kolbturner}, which yields an axion density of this order.

To think about the axion momentum distribution, we have first stressed that axion strings, due to their infrared divergent tension, need to be carefully defined. In particular, we have focused on a low energy effective action consisting of axions with momenta below some cutoff, $k_0$, and string solutions defined by a cutoff in space of order $k_0^{-1}$.  
By studying the system in the conformal frame, we have seen that, generically, the assumption that at the time of
the Peccei-Quinn phase transition, $\langle (\vec \nabla a)^2 \rangle \sim f_a^2 H^2$ leads to a distribution of axions for momenta greater than $H$ of the form of \ref{logdistribution}, with $q=1$ and an upper cutoff on $k$ of order $k_0$ (this is consistent with the most recent simulation \cite{Buschmann:2021sdq}). The string, defined as above, gives an energy distribution of the form of Eq.(\ref{logdistribution}), matching onto the low energy axion contribution provided there is of order one string per Hubble
volume.

Indeed, one of the main points of this paper has been the study of this effective action for string modes and low momentum axions.  This is particularly straightforward in a non-relativistic limit, where it is easy to identify the collective coordinates of the string solution and to compute their couplings to low momentum axions.  One striking feature of this action is that the axion-string coupling is not given by the Kalb-Ramond action, which has frequently been used in the study of radiation from axion cosmic strings.

In this framework, we have argued that $q$ is not greater than one, for reasons in addition to the compactness question alluded to above.  In particular, for non-relativistic motion, the axion-string coupling suppresses the production of low momentum axions.  Such a suppression can also be understood if string motion is initially slow, so that an adiabatic approximation is appropriate.  If the motion is fast, a sudden approximation is appropriate, and this fixes $q=1$.  All of this is consistent with a view that {\it global} cosmic strings, with their divergent tension, play no special role in the production of axion dark matter; they are simply part of a ``soup" of phenomena involving Hubble scale variation.   

This leaves us, then, with several sources of order one uncertainty in the axion dark matter density.  These include
 \begin{enumerate}
 \item  Limited knowledge of the QCD free energy in the relevant region.
 \item  Imperfect knowledge of the low momentum axion distribution.
 \end{enumerate}
The first item requires improved lattice simulations of the topological susceptibility
in finite temperature QCD.
In earlier work, it has been noted that the axion density is not terribly sensitive to the first item above\cite{dinedraper}.  It is, more specifically, only mildly sensitive to interpolations between the known low and high temperature behaviors of the free energy.  The second uncertainty, which again has only mild effects
of the connection of the dark matter density to the axion mass, is also a worthwhile subject of simulations.

\vskip 1cm
\noindent
\noindent
{\bf Acknowledgements:}  This work was supported in part by the U.S. Department of Energy grant number DE-FG02-04ER41286. NF was partly supported by the DoE Early Career Grant DE-SC0017840. AG was partly supported by NSF CAREER grant PHY-1915852 and by the U.S. Department of Energy under grant No. DE- SC0007914. We thank Marco Gorghetto for helpful conversations. 

\appendix\label{sec:NGKReffectiveAction}
\section{Kalb-Ramond effective action for global strings}
An effective description for global strings is given by the Kalb-Ramond action 
\beq
\label{eq:NGKR}
S = -\mu \int d\,\tau d\,\sigma \sqrt{- \gamma}\, - 2 \pi f_{a} \int  d\,\tau d\,\sigma B_{\mu \nu} \partial_{\alpha}  X^{\mu} \partial_{\beta} X^{\nu} \epsilon^{\alpha\beta} \, - \frac{1}{6} \int \, d^{4}x H^{\mu\nu\rho}H_{\mu\nu\rho}\,,
\eeq
where, the first term is the familiar Nambu{-}Goto action for the free string and the second term represents the interaction of the string with $B_{\mu\nu}$. Here, $\mu$ is the tension of the string and $\sigma^{\alpha}=(\tau,\sigma)$ are the wordsheet coordinates. The additional terms are the induced metric on the worldsheet,
\beq
\gamma_{\alpha \beta} = \frac{\partial X^{\mu}}{\partial \sigma^{\alpha}}\frac{\partial X^{\nu}}{\partial \sigma^{\beta}}\eta_{\mu\nu} \,,
\eeq
the antisymmetric tensor, $B_{\mu \nu}$, and its field strength tensor $H_{\mu \nu \rho}$ (see Sec.\ref{Sec:Comparison}). They are related by
\beq
H_{\mu \nu \rho} = \partial_{\mu}B_{\nu \rho} +  \partial_{\nu}B_{\rho\mu} +  \partial_{\rho}B_{\mu\nu}\,.
\eeq
The equations of motions for the string and the $B$-field are obtained by varying the action with respect to $X^{\mu}$ and $B^{\mu\nu}$, respectively,
\beq
\label{eq:EOM}
&\mu \left( \ddot{X}^{\mu}- X^{\prime\prime\mu}\right) = 2\pi f_{a}H^{\mu\nu\rho}\left(\dot{X}_{\nu}X^{\prime}_{\rho} - \dot{X}_{\rho}X^{\prime}_{\nu}\right) \\
&\Box B^{\mu \nu} = 2 \pi f_{a} \int d\,\tau d\,\sigma \left(\dot{x}^{\mu} x^{\prime\nu} - \dot{x}^{\nu} x^{\prime \mu}  \right) \delta^{(4)}(x-X_0(\tau,\sigma)) \,,
\eeq
where overdots are derivatives with respect to $\tau$ and primes with respect to $\sigma$. Here, we worked in the conformal gauge, meaning $\dot{X}^{2} + X^{\prime\,2}=0$ and $\dot{X} \cdot X^{\prime}=0$. Also, we employed the Lorentz gauge for the antisymmetric tensor, $\partial_{\mu} B^{\mu\nu}=0$.

\section{Radiation by moving axion strings} 
In this appendix we present a derivation of Eq.~(\ref{eq:rad}), the formula for the radiation of axions by global strings undergoing arbitrary motion based on the Kalb-Ramond effective action Eq.~(\ref{eq:NGKR}).
The equation of motion for the $B$-field derived from the action, Eq.(\ref{eq:EOM}), is
\beq
 \Box B^{\mu \nu} = J^{\mu \nu}\,.
\eeq 
where the string source $J^{\mu\nu}(x)$ is characterized by the string coordinates $X_0^\mu(\tau,\sigma)$
\beq
 J^{\mu \nu}(x;x_0) = 2\pi f_{a} \int_{-\infty}^{\infty} \dd \tau \, \int_{-\pi}^{\pi} \dd \sigma \, \left[\dot{x}^{\mu} x^{\prime\nu} - \dot{x}^{\nu} x^{\prime \mu}  \right] \delta^{(4)}(x-X_0(\tau,\sigma))
\eeq
with $\dot{x}^\mu \equiv \partial x^\mu/\partial \tau$ and $x'^\mu \equiv \partial x^\mu/\partial\sigma$.  The causal solution to the wave equation is
\beq
B^{\mu\nu}(x) = \int \dd^{4} y \, D_{r}(x-y) J^{\mu\nu}(y)
\eeq
where $D_{r}(x-y) =  \dfrac{1}{2\pi} \Theta (x^0-y^0) \,\delta \left[ (x-y)^2\right]$ is the massless retarded Green function.  After inserting the source $J^{\mu \nu}$ and integrating over $\dd^{4}y$, fixing $y\rightarrow x_{0}$ we have
\beq
B^{\mu \nu} =f_a \int_{-\infty}^{\infty} \dd \tau \, \int_{-\pi}^{\pi} \dd \sigma \, \left[\dot{X}_{0}^{\mu} X_{0}^{\prime\nu} - \dot{X}_{0}^{\nu} X_{0}^{\prime \mu}  \right] \Theta (t-X^0_{0}) \,\delta \left( (x-X_{0})^2\right) \,.
\eeq
To construct the radiation field strength $H^{\mu \nu \rho}= \partial^{\mu}B^{\nu\rho} + \partial^{\nu}B^{\rho\mu} +\partial^{\rho}B^{\mu\nu}$, we require the gradient
 \beq \label{eq:gradB}
\partial^{\rho}B^{\mu\nu} = f_a \int \dd \tau \, \dd \sigma \, \left[\dot{X}_{0}^{\mu} X_{0}^{\prime\nu} - \dot{X}_{0}^{\nu} X_{0}^{\prime \mu}  \right] \partial^{\rho}\left[\Theta (t-X^0_{0}) \,\delta \left( (x-X_{0})^2\right) \right]
 \eeq
Since we are interested in radiation field at large times $t\gg X^0_{0}$, we may pull the Heaviside step function out of the gradient. The gradient of the delta function may then be written as  
\beq 
\partial^{\rho}\big[\delta\big((x-X_{0})^2\big)\big] =
 - \dfrac{(x-X_{0})^{\rho}}{(x-X_{0})\cdot\pd{X_{0}}{\tau}} \, \dfrac{\partial}{\partial \tau} \delta \left((x-X_{0})^2\right)\,,
\eeq
so that (\ref{eq:gradB}) becomes
\beq
\partial^{\rho} B^{\mu \nu} &= - f_a \int \dd \tau \, \dd \sigma \, \left[\dot{X}_{0}^{\mu} X_{0}^{\prime\nu} - \dot{X}_{0}^{\nu} X_{0}^{\prime \mu}  \right] \Theta (t - X^0_{0}(\sigma, \tau)) \dfrac{(x-X_{0})^{\rho}}{(x-X_{0}) \cdot \dot{X}_{0}} \, \dfrac{\partial}{\partial \tau} \delta \left((x-X_{0})^2\right)\\
&=  f_a \int \dd \tau \, \dd \sigma \, \dfrac{\partial}{\partial \tau} \Big\{\left[\dot{X}_{0}^{\mu} X_{0}^{\prime\nu} - \dot{X}_{0}^{\nu} X_{0}^{\prime \mu}  \right] \Theta (t - X^0_{0}(\sigma, \tau)) \dfrac{(x-X_{0})^{\rho}}{(x-X_{0}) \cdot \dot{X}_{0}} \Big\}  \delta \left((x-X_{0})^2\right)\,,
\eeq 
where in the second line, we have integrated $\tau$ by parts.  Then after using
\beq
\delta \left[ (x-X_{0})^2\right] = \dfrac{\delta(\tau - \tau_{R})}{2(x-X_{0}) \cdot  \dot{X}_{0} \Bigr\rvert_{\mathrlap{\tau = \tau_{R}}}}\,,
\eeq
where $\tau_R$ is the retarded tau coordinate defined by the root of the delta function, we integrate over $\tau$ and to fix $\tau\rightarrow \tau_{R}$ to obtain
\beq
\partial^{\rho} B^{\mu \nu} = - \dfrac{f_a}{2} \int \dd \sigma \, \dfrac{1}{(x-X_{0}) \cdot  \dot{X}_{0}} \, \dfrac{\partial}{\partial \tau} \left. \left[ \dfrac{(x-X_{0})^{\rho}  \left(\dot{X}_{0}^{\mu} X_{0}^{\prime\nu} - \dot{X}_{0}^{\nu} X_{0}^{\prime \mu}  \right)  }{(x-X_{0}) \cdot  \dot{X}_{0}}\right]\right\rvert_{\tau=\tau_{R}}
\eeq
Abbreviating $f_0^{\mu\nu} = \dot{X}_{0}^{\mu} X_{0}^{\prime\nu} - \dot{X}_{0}^{\nu} X_{0}^{\prime \mu}$, and computing the derivative, we get
\beq
\dfrac{\partial}{\partial \tau}[\dots] &= \underbrace{-\frac{\dot{X}_{0}^{\rho}f_{0}^{\mu \nu}}{(x-X_{0}) \cdot  \dot{X}_{0}}}_{\text{Zero by $H^{\mu\nu\rho}$ antisymm}} + \frac{(x-X_{0})^{\rho} \, \dot{f}_{0}^{\mu \nu}}{(x-X_{0}) \cdot  \dot{X}_{0}} - \dfrac{(x-X_{0})^{\rho} \, f_{0}^{\mu\nu} \left[ (-\dot{X}_{0})\cdot \dot{X}_{0} + (x-X_{0}) \cdot \ddot{X}_{0}\right]}{\left[ (x-X_{0}) \cdot  \dot{X}_{0} \right]^{2}} \\
&= \frac{(x-X_{0})^{\rho} \, f_{0}^{\mu\nu} (\dot{X}_{0})^{2}}{\left[ (x-X_{0}) \cdot  \dot{X}_{0} \right]^{2}} + \frac{(x-X_{0})^{\rho} \, \dot{f}_{0}^{\mu \nu} (x-X_{0}) \cdot  \dot{X}_{0} -(x-X_{0})^{\rho} f_{0}^{\mu \nu}(x-X_{0}) \cdot \ddot{X}_{0} }{\left[ (x-X_{0}) \cdot  \dot{X}_{0} \right]^{2}} 
\eeq
Then, we have 
\begin{align}
\nonumber \partial^{\rho} B^{\mu \nu} =&  - \frac{f_a}{2} \int \dd \sigma \, \underbrace{\frac{1}{(x-X_{0}) \cdot  \dot{X}_{0}} }_{\sim 1/|\vec{x}|}\left. \Bigg[ \underbrace{\frac{(x-X_{0})^{\rho} \, f_{0}^{\mu\nu} (\dot{X}_{0})^{2}}{\left[ (x-X_{0}) \cdot  \dot{X}_{0} \right]^{2}}}_{\text{velocity fields} \sim 1/|\vec{x}| } \right. \\
& \hspace{4cm} + \underbrace{\dfrac{(x-X_{0})^{\rho}(x-X_{0})_{\lambda}\left[ \dot{f}_{0}^{\mu \nu} \dot{X}_{0}^{\lambda} - f^{\mu \nu}_{0} \ddot{X}_{0}^{\lambda} \right]}{\left[ (x-X_{0}) \cdot  \dot{X}_{0} \right]^{2}}}_{\text{acceleration field}\sim |\vec{x}|^{0}}\Bigg] \Bigg|_{\tau=\tau_{R}}\,.
\end{align}
Dropping the velocity field contributions, the radiation fields are given by
\beq
\partial^{\rho} B^{\mu \nu}_{\mathrm{rad}} \sim - \dfrac{f_a}{2} \int \dd \sigma \, \dfrac{1}{(x-X_{0}) \cdot  \dot{X}_{0}} \left. \left[  \dfrac{(x-X_{0})^{\rho} \left[ \dot{f}_{0}^{\mu \nu} (x-X_{0}) \cdot \dot{X}_{0}- f^{\mu \nu}_{0} (x-X_{0}) \cdot \ddot{X}_{0}\right]}{\left[ (x-X_{0}) \cdot  \dot{X}_{0} \right]^{2}}\right] \right\rvert_{\tau=\tau_{R}}
\eeq
We now proceed to construct the Poynting vector defined by $T^{0 i} = H^{0ik}H^{i}_{\,jk}$.  Denoting the observation point as $x^{\mu}=(t;|\vec{x}| \hat{n})$, and employing the string coordinates
\beq
X^{\mu}_{0}&=R_{0}(\tau ; \vec{x}_{0})\\
\dot{X}^{\mu}_{0}&=R_{0}(1 ; \vec{v}_{0})\\
\ddot{X}^{\mu}_{0}&=R_{0}(0 ; \vec{a}_{0})\\
X^{\prime \mu}_{0}&=R_{0}(0 ; \vec{x}^{\, \prime}_{0})\,,
\eeq
where $\vec{x}_{0},\vec{v}_{0},\vec{a}_{0}$ are dimensionless proper position, velocity and acceleration of the segment.  In the radiation zone, $(t-R_{0}\tau_{R}) \approx |\vec{x}|$ so that
\beq
(x-X_{0})^{\mu} &\sim (|\vec{x}|, |\vec{x}| \hat{n}) \\
(x-X_{0})\cdot \dot{X}_{0} &\sim |\vec{x}|R_{0}(1-\hat{n}\cdot\vec{v})\\
(x-X_{0})\cdot \ddot{x}_{0} &\sim -|\vec{x}|R_{0}\hat{n}\cdot\vec{a}.\\
 \eeq
Then
\beq
H^{\mu \nu \rho} = - \dfrac{f_a}{2} \int \dd \sigma \, \dfrac{|\vec{x}|^{2} R_{0}}{|\vec{x}|^{3}R_{0}^{3}(1-\hat{n}\cdot\vec{v})^{3}} \left\lbrace (1,\hat{n})^{\mu} \left[ \dot{f}_{0}^{\nu\rho} \, (1-\hat{n}\cdot\vec{v})  + f^{\nu \rho} \hat{n}\cdot\vec{a} \right]  + \mathrm{cyclic}\right\rbrace
\eeq 
and
\beq
H^{0jk} = - \dfrac{f_a}{2 |\vec{x}|R_{0}^{2}} \int \dd \sigma \, \dfrac{1}{(1-\hat{n}\cdot\vec{v})^{3}} &\left\lbrace \left[ \dot{f}_{0}^{jk}+\hat{n}^{j}\dot{f}_{0}^{k0}+\hat{n}^{k}\dot{f}_{0}^{0j} \right] (1-\hat{n}\cdot\vec{v}) \right. \\
& +  \left. \left[ f_{0}^{jk}+\hat{n}^{j}f_{0}^{k0} +\hat{n}^{k}f_{0}^{0j} \right] \hat{n}\cdot\vec{a}  \right\rbrace
\eeq 
and
\beq
H^{ijk} = - \dfrac{f_a}{2 |\vec{x}|R_{0}^{2}} \int \dd \sigma \, \dfrac{1}{(1-\hat{n}\cdot\vec{v})^{3}} &\left\lbrace  \hat{n}^{i} \left[\dot{f}_{0}^{jk}(1-\hat{n}\cdot\vec{v}) + f_{0}^{jk} \hat{n}\cdot\vec{a} \right]\right. \\
& + \hat{n}^{j} \left[\dot{f}_{0}^{ki}(1-\hat{n}\cdot\vec{v}) + f_{0}^{ki} \hat{n}\cdot\vec{a} \right] \\
& + \left. \hat{n}^{k} \left[ \dot{f}_{0}^{ij}(1-\hat{n}\cdot\vec{v}) + f_{0}^{ij} \hat{n}\cdot\vec{a} \right] \right\rbrace
\eeq 
Notice that $H^{0jk}$ and $H^{ijk}$ can be written as total derivatives
\beq
H^{0jk} &= - \dfrac{f_a}{2 |\vec{x}|R_{0}^{2}} \int \dd \sigma \, \dfrac{1}{(1-\hat{n}\cdot\vec{v})} \dfrac{\partial}{\partial \tau} \left[ \dfrac{f_{0}^{jk}-\hat{n}^{j}f_{0}^{0k} +\hat{n}^{k}f_{0}^{0j}}{1-\hat{n}\cdot\vec{v}} \right]\\
H^{ijk} &= - \dfrac{f_a}{2 |\vec{x}|R_{0}^{2}} \int \dd \sigma \, \dfrac{1}{(1-\hat{n}\cdot\vec{v})} \dfrac{\partial}{\partial \tau} \left[ \dfrac{\hat{n}^{i} f_{0}^{jk}+\hat{n}^{j}f_{0}^{ki} +\hat{n}^{k}f_{0}^{ij}}{1-\hat{n}\cdot\vec{v}} \right].
\eeq
It is convenient to introduce gauge invariant vector $\vec{V}$ and pseudoscalar $P$ fields (analogous to $\vec{E}$ and $\vec{B}$) defined by:
\beq
V^{i}&=\frac{1}{2}\epsilon^{ijk}H^{0jk} \\
&=- \dfrac{f_a}{2 |\vec{x}|R_{0}^{2}} \int \dd \sigma \, \dfrac{1}{(1-\hat{n}\cdot\vec{v})} \dfrac{\partial}{\partial \tau}\left[ \dfrac{\frac{1}{2}\epsilon^{ijk}f^{jk}-\epsilon^{ijk} \hat{n}^{j}f^{0k} }{1-\hat{n}\cdot\vec{v}} \right]\\
&=- \dfrac{f_a}{2 |\vec{x}|} \int \dd \sigma \, \dfrac{1}{(1-\hat{n}\cdot\vec{v})} \dfrac{\partial}{\partial \tau}\left[ \dfrac{(\vec{v}-\hat{n})\times\vec{x}^{\,\prime}}{1-\hat{n}\cdot\vec{v}} \right]
\eeq
and similarly
\beq
P&=\frac{1}{6}\epsilon^{ijk}H^{ijk}\\
&=- \dfrac{f_a}{2 |\vec{x}|R_{0}^{2}} \int \dd \sigma \, \dfrac{1}{(1-\hat{n}\cdot\vec{v})} \dfrac{\partial}{\partial \tau}\left[ \dfrac{\frac{3}{6}  \epsilon^{ijk} \hat{n}^{i}f^{jk} }{1-\hat{n}\cdot\vec{v}} \right]\\
&=- \dfrac{f_a}{2 |\vec{x}|} \int \dd \sigma \, \dfrac{1}{(1-\hat{n}\cdot\vec{v})} \dfrac{\partial}{\partial \tau}\left[ \dfrac{\frac{3}{6} \hat{n}\cdot(\vec{v}\times\vec{x}^{\,\prime}) }{1-\hat{n}\cdot\vec{v}} \right]\\
&=\hat{n}\cdot \vec{V}
\eeq
where we use the following identities:
\beq
f^{jk}&=R_{0}^{2} (\vec{v}^{j}\vec{x}^{\, \prime k}- \vec{x}^{\, \prime j}\vec{v}^{k})\\
\epsilon^{ijk} f^{jk}&=2R_{0}^{2} (\vec{v} \times \vec{x}^{\, \prime})
\eeq
Then $T^{0i}=2V^{i} P$. To obtain the frequency spectrum perform F.T of $V^{i}$ and $P$.

\beq
\widetilde{V}^{i}(\omega) &= \int_{-\infty}^{\infty} \dd t \, e^{i \omega t} V^{i}(t, \vec{x})\\
\widetilde{P}(\omega) &= \int_{-\infty}^{\infty} \dd t \, e^{ i \omega t} P(t, \vec{x})
\eeq

 Then
 \beq
 \widetilde{V}^{i}(\omega)=- \dfrac{f_a}{2 |\vec{x}|} \int \dd \sigma \, \int_{-\infty}^{\infty} \dd t \, \dfrac{ e^{i \omega t} }{(1-\hat{n}\cdot\vec{v})} \dfrac{\partial}{\partial \tau} \left. \left[ \dfrac{(\vec{v}-\hat{n})\times\vec{x}^{\,\prime}}{1-\hat{n}\cdot\vec{v}} \right] \right\rvert_{\tau=\tau_{R}}
 \eeq
Changing $t= R_{0} (\tau_{R} - \hat{n}\cdot\vec{x}_{0})$ and $\dd t= R_{0} (1- \hat{n}\cdot\vec{v})\dd \tau$ and integrating by parts we have $\omega$
 \beq
 \widetilde{V}^{i}(\omega)=- \dfrac{i \omega f_a R_{0}^{2}}{2 |\vec{x}|} \int \dd \sigma \, \int_{-\infty}^{\infty} \dd \tau_{R} \, [(\vec{v}-\hat{n})\times\vec{x}^{\,\prime}] e^{i \omega R_{0} (\tau_{R} - \hat{n}\cdot\vec{x}_{0})} 
 \eeq
and 
 \beq
 \widetilde{P}(\omega)=- \dfrac{i \omega f_a R_{0}^{2}}{2 |\vec{x}|} \int \dd \sigma \, \int_{-\infty}^{\infty} \dd \tau_{R} \, [\hat{n} \cdot (\vec{v}\times\vec{x}^{\,\prime})] e^{i \omega R_{0} (\tau_{R} - \hat{n}\cdot\vec{x}_{0})} 
 \eeq
Under the assumption of $\vec{V}$ points in the $\hat{n}$ direction, we can obtain $|\vec{V}|=\hat{n}\cdot\vec{V}$, and recall that $P=\hat{n}\cdot\vec{V}$.  The total energy radiated is
\beq
I=\int \dd t\, \int \dd \vec{S}\cdot \vec{P} = \int \dd t\, \dd \Omega \, |\vec{x}|^{2} \hat{n}\cdot \vec{S}
\eeq

where $\vec{S}=2 \vec{V} P$, then
\beq
\dfrac{\dd I}{\dd \Omega} &= 2 |\vec{x}|^{2} \int \dd t\,  \hat{n}\cdot\vec{V} P\\
&=2 |\vec{x}|^{2} \int \dd t\,  |P|^{2}\\
\eeq
Inserting Fourier expansions
\beq
\dfrac{\dd I}{\dd \Omega} &=2 |\vec{x}|^{2} \int \dd t\, \int \dfrac{\dd \omega}{2 \pi} e^{-i \omega t} \widetilde{P}(\omega) \int \dfrac{\dd \omega^{\prime}}{2 \pi} e^{i \omega^{\prime}t} \widetilde{P}^{*}(\omega^{\prime}) 
&=2 |\vec{x}|^{2} \int \dfrac{\dd \omega}{2 \pi}  |\widetilde{P}(\omega) |^{2}
\eeq
so
\beq
\dfrac{\dd^{2}I}{\dd \omega \, \dd \Omega} = \dfrac{1}{\pi} |\widetilde{P}(\omega) |^{2} = \dfrac{\omega^{2} f_{a}^{2} R_{0}^{4}}{4 \pi} \left\lvert \int \dd \sigma \int \dd \tau_{R} \,  \hat{n} \cdot (\vec{v}\times\vec{x}^{\,\prime}) e^{i \omega R_{0}(\tau_{R} - \hat{n}\cdot\vec{x}_{0})} \right\rvert^{2}
\eeq

\bibliography{axions_cosmic_strings}{}
\bibliographystyle{utphys}

 \end{document}